\documentclass[12pt]{article}

\usepackage{amsmath,amsfonts}

\def\al{\alpha}

\def\ga{\gamma}
\def\la{\lambda}
\def\th{\theta}
\def\si{\sigma}

\newcommand{\half}{\frac{1}{2}}
\newcommand{\order}[1]{{\mathcal{O}}(#1)}
\newcommand{\betah}{\hat{\beta}}
\newcommand{\bp}{U}
\newcommand{\bm}{V}

\newcommand{\p}{p}
\newcommand{\q}{q}
\newcommand{\pt}{\hat{p}}
\newcommand{\qt}{\hat{q}}

\newcommand{\beqa}{\begin{eqnarray}}
\newcommand{\eeqa}{\end{eqnarray}}
\newcommand{\one}[1]{\stackrel{1}{#1}}
\newcommand{\two}[1]{\stackrel{2}{#1}}
\newcommand{\lterms}[1]{\raisebox{-0.23cm}{~\shortstack{ $=$ \\
${\vspace{-0.2cm} _{#1}}$}}}

\begin{document}

%
%
%

\centerline{\Large \bf Coupling integrable field theories to}
\vspace{2mm}

\centerline{\Large \bf mechanical systems at the boundary}

\vskip 1cm

\centerline{\large P. Baseilhac\,\footnote{ \tt e-mail:
pb18@york.ac.uk} \ \ and\ \ \ G.W. Delius\,\footnote{ \tt e-mail:
gwd2@york.ac.uk}}

\vspace{0.5cm}

\centerline {\it Department of Mathematics, University of York }

\centerline {\it Heslington, York YO10 5DD, United Kingdom}

\vspace{1mm}

\centerline {\small\tt http://www.york.ac.uk/mathematics/physics/}

\vspace{1cm}

\begin{abstract}
We present an integrable Hamiltonian which describes the
sinh-Gordon model on the half line coupled to a non-linear
oscillator at the boundary. We explain how we apply Sklyanin's
formalism to a dynamical reflection matrix to obtain this model.
This method can be applied to couple other integrable field
theories to dynamical systems at the boundary. We also show how to
find the dynamical solution of the quantum reflection equation
corresponding to our particular example.
\end{abstract}

\section{Introduction}

Integrable field theories in two dimensions provide us with a
theoretical laboratory to study non-perturbative phenomena in
high-energy physics. They also possess numerous applications in
fluid dynamics, statistical physics, condensed matter physics, and
quantum optics.

In recent years it has become possible to extend many of the
techniques and results of integrable models to field theories with
boundaries, i.e., theories defined either on the half line or on
an interval, see for instance
\cite{Che84,Skl87,Skl88,Gho94,Cor96}.

We introduce a new class of integrable field theories with
boundaries where, instead of imposing fixed boundary conditions,
we couple the boundary field to a mechanical system.

We begin by giving a very concrete example in section
\ref{sect:sg} where we describe the sinh-Gordon or sine-Gordon
model coupled to a non-linear oscillator at the boundary. Then in
section \ref{sect:c} we explain more generally how one can couple
an integrable field theory to a mechanical system in such a way
that the integrability is not broken. We use a generalization of
Sklyanin's technique for the construction  of integrable boundary
conditions. The new ingredient is that the solution of the
reflection equation is chosen to depend on boundary degrees of
freedom. In section \ref{sect:sgc} we specialize this technique to
the case of the sine-Gordon theory, providing some details of how
we arrived at the model described in section \ref{sect:sg}.
Finally, in section \ref{sect:q} we describe how to obtain a
particular dynamical solution of the quantum reflection equation.
This is the solution whose classical limit went into the
construction in the previous section. We end with discussions in
section \ref{sect:d}.

Before we begin let us survey other works which deal with coupling
to degrees of freedom at the boundaries. These fall into three
categories:

$\bullet$ ~Field theory: In \cite{Baz98} the sine-Gordon model is
coupled at the quantum level to a q-oscillator.
In \cite{Nep01} it was observed that in order to derive the fixed
boundary conditions for the supersymmetric sine-Gordon model from
an action, it was convenient to introduce a fermionic variable at
the boundary which could however be integrated out again
immediately. In \cite{Fue95} a free fermion field theory is
coupled to a dynamical boundary degree of freedom.

$\bullet$ ~Integrable quantum spin chains: chains coupled to extra
spins on the boundary are used in the study of Kondo impurities
coupled to strongly correlated electron systems, see e.g.
\cite{Wan97}. In \cite{Zho98,Fra98} new dynamical solutions of the
quantum reflection equation which are not of the RKR type were
used to describe the coupling of the boundary spins.

$\bullet$ ~Systems with a finite number of degrees of freedom:
Kuznetsov has used dynamical solutions of the classical reflection
equation to couple mechanical tops to integrable non-linear
lattices such as the Heisenberg chain or Toda lattices. Our work
can be seen as the extension of these ideas to field theory while
preserving integrability.

\section{The sinh-Gordon model coupled to an oscillator at the boundary\label{sect:sg}}

In this section we present an integrable Hamiltonian describing
the coupling of the sinh-Gordon field theory to a non-linear
oscillator at the boundary. We leave the details of how we
constructed this Hamiltonian for later sections.

The sinh-Gordon model describes a relativistic 1+1 dimensional
self-interacting massive bosonic field $\phi(x,t)$. The
Hamiltonian of the sinh-Gordon model restricted to the half-line
is
\begin{equation}\label{hshg}
  H_{\text{shG}}=\int_{-\infty}^0 dx\left(
  \half\pi^2+\half(\partial_x\phi)^2+
  \frac{m^2}{\betah^2}(\cosh \betah\phi -1)\right).
\end{equation}
Here $\pi$ is the conjugate momentum to the field $\phi$, i.e.,
\begin{equation}
  \{\pi(x),\phi(x)\}=\delta(x-y)\,.
\end{equation}
$\betah$ is the real sinh-Gordon coupling constant and $m$ sets
the mass scale. We let space be the half line from $x=-\infty$ to
$x=0$. As usual the field $\phi$ is assumed to vanish at
$x=-\infty$ but not at $x=0$. As will become clear from the
construction in section \ref{sect:c} we could also have taken
space to be an interval $[x_-,x_+]$ and placed a mechanical system
at both ends.

We describe the coupling of the sinh-Gordon field $\phi(0)$ at the
boundary $x=0$ to a non-linear oscillator through the boundary
Hamiltonian
\begin{equation}\label{hosc}
  H_{\text{osc}}=\frac{2m}{\betah^2}\left(
  \cosh\left(\frac{\betah}{\sqrt{2M m}}\,\p\right)e^{-\betah\phi(0)/2}+
  \cosh\left(\frac{\betah\sqrt{Mm}}{2\sqrt{2}}\,\q\right)e^{\betah\phi(0)/2}
  \right).
\end{equation}
Here $q$ and $p$ are the position and momentum variables of the
oscillator and they obey the canonical Poisson bracket relation
$\{p,q\}=1$. The new free parameter $M$ determines the mass of the
oscillator.

It is instructive to expand the Hamiltonian for the oscillator for
small $p$ and $q$. One obtains
\begin{equation}\label{hosce}
  H_{\text{osc}}=\frac{1}{2M_\phi}\p^2+\frac{M_\phi}{2}\omega^2 \q^2+
  \frac{4m}{\betah^2}\cosh\left(\frac{\betah\phi(0)}{2}\right)+
  \order{\p^4}+\order{\q^4}\,,
\end{equation}
where
\begin{equation}\label{mphi}
  M_\phi=M e^{\betah\phi(0)/2}~~~~\text{and}~~~~\omega=m/2\,.
\end{equation}
We see that the frequency $\omega$ of the oscillator is fixed by
the requirement of integrability to be equal to half the mass of
the sinh-Gordon field. The effective mass of the oscillator
depends on the value of the sinh-Gordon field at the boundary. The
exact form of the higher non-linear terms in the Hamiltonian of
the oscillator are fixed by integrability.

To shorten the formulas we introduce the rescaled variables
\begin{equation}
  \pt=\frac{\betah}{\sqrt{2M m}}\p~~~~\text{and}~~~~
  \qt=\frac{\betah\sqrt{Mm}}{2\sqrt{2}}\q
\end{equation}
which have Poisson bracket $\{\pt,\qt\}=\betah^2/4$. In terms of
these the boundary equations of motion are
\begin{align}\label{peom}
  \frac{d}{dt}\pt&=\{H_{\text{osc}},\pt\}=-\frac{m}{2}\,
  e^{\betah\phi(0)/2} \sinh \qt\,,\\
  \frac{d}{dt}\qt&=\{H_{\text{osc}},\qt\}=\frac{m}{2}\,
  e^{-\betah\phi(0)/2} \sinh \pt\,.\label{qeom}
\end{align}

To determine the equations of motion of the sinh-Gordon field
$\phi$ one needs to use the full Hamiltonian
\begin{equation}
  H=H_{\text{shG}}+H_{\text{osc}}.
\end{equation}
One finds
\begin{align}\label{phieom}
  \frac{d}{dt}\phi(x)=&\{H,\phi(x)\}=\pi(x),\\
  \label{pieom}
  \frac{d}{dt}\pi(x)=&\{H,\pi(x)\}=\partial_x^2\phi(x)-
  \frac{m^2}{\betah}\sinh\betah\phi(x)\\
  &-\delta(x)\left(\partial_x\phi(0)-\frac{m}{\betah}\left(
  e^{-\betah\phi(0)/2}\cosh\pt-e^{\betah\phi(0)/2}\cosh\qt\right)\right).
  \nonumber
\end{align}

Note the term proportional to $\delta(x)$ in the equation of
motion for $\pi(x)$. It has two sources: the $\partial_x\phi$
arises when one performs a partial integration in the sinh-Gordon
Hamiltonian which produces a boundary contribution at $x=0$ and
the other two terms arise because the Hamiltonian of the
oscillator contains $\phi(0)$.

We are only going to allow solutions to these equations which are
continuous on the left half-line $[-\infty,0]$. So in particular
we require continuity of $\pi(x)$ at $x=0$. This implies that the
$\delta(x)$ term in the equation of motion for $\pi(x)$ vanishes
because this term would otherwise force $\pi(x)$ to develop a
discontinuity at $x=0$. We therefore know that the solutions will
satisfy
\begin{equation}\label{bc}
  \partial_x\phi(0)=\frac{m}{\betah}\left(
  \cosh\pt\ e^{-\betah\phi(0)/2}-\cosh\qt\ e^{\betah\phi(0)/2}\right).
\end{equation}

We can combine the equations of motion \eqref{phieom} and
\eqref{pieom} to obtain the usual sinh-Gordon equation of motion
\begin{equation}\label{seom}
  \partial_t^2\phi-\partial_x^2\phi=
  \frac{m^2}{\betah}\sinh\betah\phi.
\end{equation}
This equation for $\phi$ together with the boundary condition
\eqref{bc} and the equations of motion \eqref{peom} and
\eqref{qeom} for the oscillator form one system of coupled
equations involving both ODE's and a PDE which needs to be solved
to obtain the time evolution of the system.

We call this system ``integrable'' because, as we will show in
section \ref{sect:sgc}, we can construct an infinite number of
conserved higher spin charges $I_n$ which are in involution with
each other. The first non-trivial charge beyond the Hamiltonian is
the spin 3 charge
\begin{align}\label{i3}
  I_3&=\int_{-\infty}^0\left(
    \frac{\betah^4}{16m^3}\left(\pi^4+6\pi^2(\partial_x\phi)^2+
    (\partial_x\phi)^4\right)
    +\frac{\betah^2}{m^3}\left((\partial_x\pi)^2+(\partial_x^2\phi)^2\right)
    \right.\nonumber\\
    &~~~~~~~~~~~~\left.
    +\frac{\betah^2}{4m}\left(\pi^2+5(\partial_x\phi)^2\right)\cosh\betah\phi
    +\frac{m}{8}\left(\cosh 2\betah\phi - 1\right)
    \right)dx\nonumber\\
    &~~~+e^{3\betah\phi(0)/2}\left(\half\cosh\qt
    +\frac{1}{6}\cosh^3\qt\right)\\
    &~~~-e^{\betah\phi(0)/2}\left(\frac{3}{2}\cosh\pt
    -\half\cosh^2\qt\cosh\pt
    -\frac{\betah^2}{2m^2}\pi^2\cosh\qt\right)\nonumber\\
    &~~~+e^{-3\betah\phi(0)/2}\left(\half\cosh\pt
    +\frac{1}{6}\cosh^3\pt\right)\nonumber\\
    &~~~-e^{-\betah\phi(0)/2}\left(\frac{3}{2}\cosh\qt
    -\half\cosh^2\pt\cosh\qt
    -\frac{\betah^2}{2m^2}\pi^2\cosh\pt\right)\nonumber\\
    &~~~+\frac{2\betah}{m}\sinh\qt\,\sinh\pt\,.\nonumber
\end{align}

The boundary condition \eqref{bc} is similar in form to the
previously known integrable boundary conditions \cite{Gho94,mac}
\begin{equation}\label{obc}
  \partial_x\phi(0)=\frac{m}{\betah}\left(
  \epsilon_0\,e^{-\betah\phi(0)/2}-\epsilon_1\,e^{\betah\phi(0)/2}\right),
\end{equation}
with the crucial difference of course that the parameters
$\epsilon_0$ and $\epsilon_1$ were fixed numbers rather than
dynamical variables as in our case. Only in one case do our
boundary reduce to the fixed boundary conditions \eqref{obc},
namely when the boundary oscillator is at rest at $q=p=0$. This
corresponds to the boundary conditions with
$\epsilon_0=\epsilon_1=1$. The quantum fluctuations of the
boundary oscillator will probably imply that our model never
reduces to the usual boundary conditions in the quantum case.

After quantization the sinh-Gordon model describes scalar massive
particles. The direct solution of the quantum theory is rather
difficult \cite{Skl89}. However because of the existence of higher
spin conserved charges one knows that there is no particle
production and that the particle scattering factorizes into a
product of two-particle scattering processes. The corresponding
scattering amplitude has been obtained by analytical continuation
in the coupling constant from the breather scattering amplitude in
the sine-Gordon model \cite{Zam79}.

To describe the sinh-Gordon particles on the half-line one also
needs to give the reflection amplitudes. In the case of the fixed
boundary conditions \eqref{obc} this amplitude can again be
obtained from the corresponding breather reflection amplitude in
the sine-Gordon model \cite{Gho93}. Because we expect that the
results for our model can be obtained similarly we now turn our
attention to the sine-Gordon model.

When we let the sinh-Gordon coupling constant $\betah$ become
purely imaginary, i.e., if we set $\betah=i\beta$ with $\beta$
purely real, then the sinh-Gordon Hamiltonian \eqref{hshg} turns
into the sine-Gordon Hamiltonian
\begin{equation}\label{hsg}
  H_{\text{sG}}=\int_{-\infty}^0 dx\left(
  \half\pi^2+\half(\partial_x\phi)^2-
  \frac{m^2}{\beta^2}(\cos\beta\phi -1)\right).
\end{equation}
Under the same replacement the Hamiltonian for the oscillator
becomes
\begin{equation}\label{hoscs}
  H_{\text{osc}}=-\frac{2m}{\beta^2}\left(
  \cos\left(\frac{\beta}{\sqrt{2M m}}\,p\right)\,e^{-i\beta\phi(0)/2}+
  \cos\left(\frac{\beta\sqrt{Mm}}{2\sqrt{2}}\,q\right)\,e^{i\beta\phi(0)/2}
  \right).
\end{equation}
Notice that in general this Hamiltonian is not real. We are
familiar with this situation from imaginary coupled affine Toda
theories. There the non-real Hamiltonian implies that the
classical soliton solutions are complex. The energy of these
configurations is nevertheless real \cite{Hol92}. Similarly here
the classical boundary solutions will be complex but we expect
that their energies will be real.


\section{Sklyanin's formalism\label{sect:c}}

In a seminal two-page paper \cite{Skl87} Sklyanin described how
one can impose boundary conditions on an integrable field theory
without breaking integrability. Below we will describe how
Sklyanin's formalism  allows us to couple an integrable field
theory to a mechanical system at the boundary rather than imposing
a fixed boundary condition. This section will be easier to
understand for readers who are familiar with the approach to
integrable models described in \cite{Fad87}.

In the following we assume that there exists a pair of matrix
valued functions $a_x(\la,x)$ and $a_t(\la,x)$ which depend on the
fields of the theory, their conjugate momenta, and on a spectral
parameter $\la=e^\theta \in {\mathbb C}$, so that the classical
equations of motion of the field theory are equivalent to the Lax
pair equation
\beqa \big[\partial_x-a_x(\la,x),\partial_t-a_t(\la,x)\big]=0\ \ \
\ \mbox{for all} \ \ \ \la\label{compat}. \eeqa
Here $a_x(\la,x)$\  and \ $a_t(\la,x)$ \ depend on $x$ and $t$
only implicitly through their dependence on the fields.
$\partial_x$\ and\  $\partial_t$\  denote total differentiation
with respect to the space or time variable. If\  $a_x(\la,x)$\ and
\ $a_t(\la,x)$ \ are thought of as the components of a connection
then eq. (\ref{compat}) is the zero curvature condition for this
connection.

Equation \eqref{compat} is the compatibility condition for the
overdetermined system of equations
\beqa
\frac{\partial T}{\partial x_+} &=&a_x(\la,x_+)T,\label{diffe} \\
\frac{\partial T}{\partial t}
&=&a_t(\la,x_+)T-Ta_t(\la,x_-),\nonumber \eeqa
where the transition matrix $T\equiv T(x_+,x_-,\la)$ is defined to
be a solution of the differential equations (\ref{diffe}) with the
initial conditions $T(x_-,x_-,\la)=I$. It can be expressed as the
path ordered exponential of $a_x(\la,x)$ from $x_-$ to $x_+$
\beqa
  T(x_+,x_-,\la)={\cal P}\exp\big(\int_{x_-}^{x_+}a_x(\la,x)dx\big)\label{trans}
\eeqa
so that the operators at points nearer to $x_+$ are further to the
left. We assume that the Poisson brackets for the functions
$a_x(\la,x)$ can be written in the form
\beqa \{{\one a}_x(\la_1,x),{\two a}_x(\la_2,y)\}=\delta(x-y)
\big[r(\ln(\la_1/\la_2)),{\one a}_x(\la_1,x)+{\two
a}_x(\la_2,y)\big]. \eeqa
Here we used the short hand notations ${\one
a}(\la,x)=a(\la,x)\otimes I$,
 ${\two a}(\la,x)=I \otimes a(\la,x)$ and the $r$-matrix is independent of the field
or its conjugate momentum. It follows that
\beqa
 \{{\one T}(x_+,x_-,\la_1),{\two T}(x_+,x_-,\la_2)\} =
\big[r(\ln(\la_1/\la_2)),{\one T}(x_+,x_-,\la_1){\two
T}(x_+,x_-,\la_2)\big].\label{rTT} \eeqa
%
For simplicity we assume below that the $r$-matrix has the
property \footnote{This could however be relaxed, see for example
\cite{Del98b} for an application of Sklyanin's formalism to affine
Toda theory where the $r$-matrix does not possess this property.}
\beqa r(\th)=-r(-\th)\label{rmatinv}. \eeqa
Let us now introduce two matrix valued functions $K_{\pm}(\th)$ of
the spectral parameter $\la=e^\th$, not depending on the fields.
However, in departure from the situation described by Sklyanin in
\cite{Skl87}, we let these boundary $K$ matrices be dynamical.
This means that we enlarge the phase space of the theory by
introducing extra degrees of freedom which we think of as
describing mechanical systems placed at the boundaries. We then
let the $K$ matrices depend on these new dynamical variables in
such a way that their Poisson brackets are given by the following
classical reflection equation:
\beqa \{{\one K}_{\pm}(\th),{\two K}_{\pm}(\th')\} &=&
\big[r(\th-\th'),{\one K}_{\pm}(\th){\two K}_{\pm}(\th')\big] \label{KK} \\
&& +\  {\one K}_{\pm}(\th)r(\th+\th'){\two K}_{\pm}(\th') \ -\
{\two K}_{\pm}(\th')r(\th+\th'){\one K}_{\pm}(\th).\nonumber \eeqa
Furthermore we assume that the dynamical systems on the left and
right boundary are independent so that $\{{\one
K}_{\pm}(\th),{\two K}_{\mp}(\th')\}=0$.

Following Sklyanin \cite{Skl87} we define the functional
\beqa {\cal T}(x_+,x_-,\la)=
T(x_+,x_-,\la)K_{-}(\ln\la)T^{-1}(x_+,x_-,1/\la)\ \label{Tbound}
\eeqa
which generalizes the transition matrix to the boundary case.
Using eq. (\ref{rmatinv}) and that the Poisson bracket is
antisymmetric, satisfies the Jacobi identity and has the property
$\{A,BC\}=\{A,B\}C+B\{A,C\}$ it is straightforward to show that
${\cal T}(x_+,x_-,\la)$ obeys a Poisson bracket relation similar
to those for the $K$'s given in eq. (\ref{KK}).
%
%
A family of transfer matrices is defined by
\beqa \tau(\la)=tr\big(K_{+}(\ln(\la)){\cal T}(x_+,x_-,\la)\big).
\label{tau1} \eeqa
These are in involution for any values of the spectral parameters
\beqa \{\tau(\la_1),\tau(\la_2)\}=0 \ \ \ \ \mbox{for all} \ \ \
(\la_1,\la_2) \in {\mathbb C}\ \label{invol} \eeqa
provided appropriate boundary conditions are imposed at $x_\pm$.
As the generating function $\tau(\la)$ can be expanded about the
singularities in the transition matrix $T(x_+,x_-,\la)$ it gives
an infinite number of quantities $I_n$ in involution with each
other. Among these we will identify the Hamiltonian of the model.
Then it follows that the $\{I_n\}$ are time conserved. The model
is thus integrable on the interval $[ x_-,x_+]$. Different
dynamical systems can be coupled at the two boundaries by choosing
different $K_-$ and $K_+$.

\section{Derivation of the sine-Gordon example\label{sect:sgc}}
In this section we will provide some details needed to apply the
general method described in the previous section to the
sine-Gordon model in order to derive the coupling to a boundary
oscillator described in section \ref{sect:sg}.

The equation of motion (\ref{seom}) for the sine-Gordon field is
representable as the zero-curvature condition for the Lax
connection $a_\mu(\la,x)$ written in terms of the standard Pauli
matrices $\sigma_k$, \ $k=1,2,3$, as
\beqa a_x(\la,x)&=&\frac{\beta}{4i}\frac{\partial \phi}{\partial
t}\sigma_3+
\frac{m}{4i}\big(\la+\frac{1}{\la}\big)\sin\big(\frac{\beta\phi}{2}\big)\sigma_1+
\frac{m}{4i}\big(\la-\frac{1}{\la}\big)\cos\big(\frac{\beta\phi}{2}\big)
\sigma_2, \\
a_t(\la,x)&=&\frac{\beta}{4i}\frac{\partial\phi}{\partial
x}\sigma_3+
\frac{m}{4i}\big(\la-\frac{1}{\la}\big)\sin\big(\frac{\beta\phi}{2}\big)\sigma_1+
\frac{m}{4i}\big(\la+\frac{1}{\la}\big)\cos\big(\frac{\beta\phi}{2}\big)
\sigma_2.\eeqa
We define $\pi(x,t)=\frac{\partial \phi(x,t)}{\partial t}$. Notice
that the matrix $a_x(\la,x)$ possesses two singularities located
at $|\la|=0$ and $|\la|=\infty$.

To extend the standard definition of the Poisson bracket, we
change the range of the integrals from $[-\infty,+\infty]$ to
$[x_-,x_+]$. Then, using the notation $\pi(x)\equiv\pi(x,t)$ and
$\phi(x)\equiv\phi(x,t)$ at fixed time $t$ we define the Poisson
bracket as
\beqa \{{\cal O}_1,{\cal
O}_2\}=\int_{x_-}^{x_+}\!dx\Big[\frac{\delta{\cal
O}_1}{\delta\pi(x)}
 \frac{\delta{\cal O}_2}{\delta\phi(x)}
- \frac{\delta{\cal O}_1}{\delta\phi(x)}
 \frac{\delta{\cal O}_2}{\delta\pi(x)}\Big] +  \frac{\partial {\cal O}_1}{\partial
p}\frac{\partial {\cal O}_2}{\partial q} - \frac{\partial {\cal
O}_1}{\partial q}\frac{\partial {\cal O}_2}{\partial
p}\label{def1} \eeqa
for any observable \ ${\cal O}_j$. Obviously, this bracket
possesses the basic properties of a Poisson bracket. It is skew
symmetric and satisfies the Jacobi identity. Also, we have \
$\{{\cal O}_1,{\cal O}_2{\cal O}_3\}=\{{\cal O}_1,{\cal
O}_2\}{\cal O}_3 + {\cal O}_2\{{\cal O}_1,{\cal O}_3\}$. Using the
definition (\ref{def1}), the non-vanishing Poisson brackets in the
sine-Gordon field theory at constant time slices are
$\{\pi(x),\phi(y)\}=\delta(x-y)$ and for the boundary variables
$p(t)$, $q(t)$ we have $\{p,q\}=1$.
Now, by calculating the Poisson brackets of the entries of the
matrix $a_x(\la,x)$ it gives a unique solution to eq. (\ref{rTT}):
\beqa r_{SG}(\th)=
\frac{\beta^2\cosh(\th)}{16\sinh(\th)}\big(I\otimes I - \si_{3}
\otimes\si_{3}\big) - \frac{\beta^2}{16\sinh(\th)}
\big(\si_1\otimes \si_1 + \si_2 \otimes\si_2\big). \eeqa
To construct the generating function (\ref{tau1}) of the integrals
of motion for the sine-Gordon model with dynamical boundaries, we
need $(p,q)$-dependent solutions $K_{\pm}(\th)$ to the classical
reflection equation (\ref{KK}). Let us introduce the rescaled
variables ${\tilde p}=p\beta/2{\sqrt 2}$ and ${\tilde
q}=q\beta/2{\sqrt 2}$. Then the matrix \cite{Sur90}
\begin{equation}\label{solK}
K_{+}(\th) = 2
\begin{pmatrix}
\cosh({\tilde p}+{\tilde q})e^{\th}
-\cosh({\tilde p}-{\tilde q})e^{-\th} & 2\sinh^2(\th)-2\sinh^2({\tilde p}) \\
2\sinh^2({\tilde q})-2\sinh^2(\th) & \cosh({\tilde p}-{\tilde
q})e^{\th}
-\cosh({\tilde p}+{\tilde q})e^{-\th}\\
\end{pmatrix}
\end{equation}
satisfies the classical reflection equation (\ref{KK}) with the
classical $r$-matrix $r(\th)=r_{SG}(\th)$ defined above. We could
place additional degrees of freedom at the left boundary as well
and use a $K_-$ of a similar form. However we would not learn
anything new and for simplicity we choose $K_{-}(\th)=I$ which is
a trivial solution of the classical reflection equation \eqref{KK}
and then move the left boundary off to $-\infty$, i.e., we
restrict the sine-Gordon field theory to the half-line. We then
assume that the field and its conjugate momentum satisfy the
Schwartz boundary condition $\phi(x_-,t)=0$ and $\pi(x_-,t)=0$ at
$x_-=-\infty$.  Using eq. (\ref{Tbound}) the generating function
(\ref{tau1}) becomes
\beqa
\tau(\la)=tr\big(K_{+}(\ln(\la))T(0,-\infty,\la)T(-\infty,0,1/\la)\big).\label{tau2}
\eeqa

Fixing ${\cal I}m(\th)=\frac{\pi}{2}$, the transition matrix
$T(-\infty,0,\la)$ has singularities located at $\la=\pm i\infty$
and $\la=\pm i 0$. There will exist two infinities of involutive
integrals coming from the coefficients of the Laurent expansions
about these two values. First, let us consider the asymptotic
expansion of the transfer matrix $\tau(\lambda)$ as
$|\la|\rightarrow\infty$. Substituting the solution $K_+(\th)$ of
(\ref{solK}) into (\ref{tau2}) and expanding  about $\la=i\infty$,
we obtain a Laurent series in $\la$ which provides an infinite
number of quantities $I_n$:
\beqa \ln\big(\tau(\la)/\la^2\big)=\sum_{n=-1}^{\infty}
\frac{I_{n}}{\la^n}\,.\label{dev} \eeqa
In performing this expansion it has helped us to look at how the
corresponding calculation was performed in \cite{mac} for the case
of non-dynamical $K$. We find
$I_{-1}=-\frac{imL}{2}|_{L\rightarrow\infty}$ \ and $I_0=0$. The
next quantity $I_1=-\frac{i\beta^2}{2m} H+\text{const.}$ gives
 the Hamiltonian $H$ for the system, where
\beqa H&=&\int_{-\infty}^0 dx
\Big[\Theta(-x)\Big(\frac{1}{2}\big[\pi^2+(\partial_x\phi)^2\big]
-\frac{m^2}{\beta^2}(\cos(\beta\phi)-1)\Big)\label{H}\\
&&\qquad \ \ \ \ \ +\ \delta(x)\frac{2m}{\beta^2}\left(
\cosh(\tilde{p}+\tilde{q})\ e^{i\beta\phi/2}
+\cosh(\tilde{p}-\tilde{q})\
e^{-i\beta\phi/2}\right)\Big].\nonumber \eeqa
This can be seen to agree with the Hamiltonian given in section
\ref{sect:sg} after we perform the canonical transformation
\begin{equation}\label{ct}
  \tilde{p}+\tilde{q}\rightarrow\qt\,,~~~
  \tilde{p}-\tilde{q}\rightarrow-\pt\,,~~~
  \phi(x)\rightarrow\phi(x)+2\pi/\beta\,.
\end{equation}
The next term in the expansion (\ref{dev}) of order $O(1/\la^2)$
reduces to $I_2=0$ after using the boundary condition \eqref{bc}.
Similarly the $O(1/\la^3)$ term in (\ref{dev}) reproduces the spin
3 charge $I_3$ given in \eqref{i3}, again after the canonical
transformation \eqref{ct}.
%
It is believed that the existence of
such higher local integrals of motion is a sufficient condition
for the classical
 integrability of the system.

Finally we observe that because the transfer matrix
$\tau(\lambda)$ in \eqref{tau2} is invariant under the
simultaneous replacements $\la\rightarrow -1/\la,
\phi(x,t)\rightarrow -\phi(x,t)$ and $q\rightarrow -q$, the
expansion of $\tau(\lambda)$ around the singularity at $0$  gives
analogous quantities $I_n$. Notice that $H$ and $I_3$ are
invariant under this transformation.

It is rather interesting to notice the following: If $p$ and $q$
are fixed $c$-numbers then, the $(p,q)$-dependent part of the
Poisson bracket (\ref{def1}) disappears and the l.h.s. of eq.
(\ref{KK}) vanishes. Considering the two leading terms in the
expansion in $\la$ of the classical reflection matrix $K_+(\th)$
given in eq. (\ref{solK}) we obtain
\begin{gather}
K_+(\ln(\la))\ \ \ \lterms{{|\la|}\rightarrow \infty}\ \ \ \la^2
\begin{pmatrix}
\frac{2}{\la}\cosh({\tilde p}+{\tilde q}) & 1\\
-1 &  \frac{2}{\la}\cosh({\tilde p}-{\tilde q})
\end{pmatrix}\ \ \ + \ \ O(\la^0)\label{Kmac}\,
\end{gather}
which can be shown to satisfy eq. (\ref{KK}). Let us now introduce
the notation
\beqa
&&\cosh({\tilde p}+{\tilde q})=P+iQ\,,\label{not} \\
&&\cosh({\tilde p}-{\tilde q})=P-iQ\,.\nonumber \eeqa
Multiplying $K_{+}(\ln(\la))$ by $1/\la^2$ the reflection matrix
now writes
\beqa K_{+}(\ln(\la))\ \sim \ \frac{2P}{\la}I + i\si_2 +
\frac{2iQ}{\la}\si_3\ \  + \ \ O(1/\la^2)\,. \eeqa
Up to order $1/\lambda^2$ this is the same K matrix used in
\cite{mac} to construct the integrable boundary conditions used by
Ghoshal and Zamolodchikov in \cite{Gho94}. Of course, our next
quantities $I_2$ and $I_3$  differ from the ones in \cite{mac} due
to the contributions of order $O(1/\la^2)$ that appear in our
case.

\section{Dynamical solutions of the quantum reflection equation\label{sect:q}}

In this section we look for operator valued solutions ${\cal
K}_\pm(\th;\al)$ of the quantum reflection equations \cite{Skl88}
\footnote{Our ${\cal K}_\pm$ are related to the $K_\pm$ of
\cite{Skl88} by ${\cal K}_\pm(\th+\alpha/2)=K_\pm(\th)$ and
$\alpha=\eta$.}
\begin{multline}\label{qKK-}
R(\th-\th'){\one {\cal K}}_{-}(\th+\al/2;\al)R(\th+\th'){\two
{\cal K}}_{-}(\th'+\al/2;\al)
\\
={\two{\cal K}}_{-}(\th'+\al/2;\al)R(\th+\th'){\one {\cal
K}}_{-}(\th+\al/2;\al)R(\th-\th')\,,
\end{multline}

\vspace{-12mm}

\begin{multline}\label{qKK+}
R(-\th+\th'){\one{{\cal
K}_+^t}}(\th-\al/2;\al)R(-\th-\th'){\two{{\cal
K}_+^t}}(\th'-\al/2;\al)
\\
={\two{{\cal K}_+^t}}(\th'-\al/2;\al)R(-\th-\th'){\one{{\cal
K}_+^t}}(\th-\al/2;\al)R(-\th+\th')\,.
\end{multline}
Here $R(\theta)$ is a given quantum $R$-matrix.
We see that the quantum
reflection matrix ${\cal K}_-(\th;\al)$ for the left boundary has
to satisfy a different equation to the quantum reflection matrix
${\cal K}_+(\th;\al)$ for the right boundary. The classical
reflection equation (\ref{KK}) can be obtained as a limiting case
of either of the quantum reflection equations \eqref{qKK-} and
\eqref{qKK+}\footnote{In the limit $\al\rightarrow 0$ one recovers
the classical $r$-matrix (up to an overall $\theta$-dependent
coefficient) \ \ $R(\th)\ \sim \ { I} + \al r(\th) + O(\al^2)$\ \
where $I$ is the identity matrix. The classical reflection
matrices $K_{\pm}(\th)$ are defined to be the first term in the
expansion of the quantum reflection matrices: ${\cal
K}_{\pm}(\th;\al)\! \lterms{{\al}\rightarrow 0}{K}_{\pm}(\th)\ +\
\al\delta {K}_{\pm}(\th)\ +\ O(\al^2)$.}.

One useful way to think of the reflection equation \eqref{qKK-} is
to view the entries of the matrix ${\cal K}_-(\th;\al)$ as the
generators of an associative algebra with quadratic algebra
relations given by \eqref{qKK-}. Such an algebra is often called a
reflection equation algebra. Finding a solution of a reflection
equation is therefore the same as finding a representation of the
corresponding reflection equation algebra. Below in \eqref{solqK-}
we will give an infinite dimensional representation in terms of
position and momentum operators $q$ and $p$.

Operator valued reflection matrices have several applications: 1)
they can be used in integrable lattice models or quantum spin
chains to couple additional boundary spins to the model
\cite{Fra98,Zho98}, 2) they can describe the reflection of
particle excitations from a boundary if there are degeneracies in
the boundary spectrum, and 3) their classical limit can be used to
construct integrable models \cite{Sur90,Kuz95}.

As an example, let us consider  the following trigonometric and
hyperbolic minimal solutions of the quantum Yang-Baxter equation:
\beqa R(\th,\ga)&=&\frac{a(\th)}{2}\big(I\otimes I + \si_{3}
\otimes\si_{3}\big) + \frac{b(\th)}{2} \big(I\otimes I - \si_{3}
\otimes\si_{3}\big)\nonumber\\&&+ \frac{c(\th)}{2}
\big(\si_1\otimes \si_1 + \si_2 \otimes\si_2\big). \eeqa
This $R$-matrix gives the Boltzmann weights of the six vertex
model which is known to be related to the XXZ and the XXX (in its
rational limit) spin chains. It possesses three different regimes
called
antiferroelectric (I), trigonometric (II) and ferroelectric (III)
with
\begin{align*}
  \text{(I)}~~~~~a(\th)&=\sinh(\ga-\th)
 &b(\th)&=\sinh(\th) ,
 &c(\th)&=\sinh(\ga) ,  &\ga>\th>0  ,\\
\text{(II)}~~~~~a(\th)&=\sin(\ga-\th) , &b(\th)&=\sin(\th)  ,
 &c(\th)&=\sin(\ga) ,  &\pi>\ga>\th>0 ,\\
\text{(III)}~~~~~a(\th)&=\sinh(\th+\ga)  , &b(\th)&=\sinh(\th)  ,
 &c(\th)&=\sinh(\ga) ,  &\th>0 ,  \ga>0 ,
\end{align*}
where the parameter $\ga$ characterizes the anisotropy of the
model. In particular, the trigonometric regime of the six-vertex
model (II) describes the critical (zero gap) limit of the
eight-vertex model \cite{Bax}.

A dynamic solution of the quantum reflection equation
corresponding to this $R$-matrix was already given in
\cite{Kuz94,Kuz96} without derivation. We will derive a similar
solution below but with some minor quantum adjustments.

For further convenience, defining the $K$-matrix at the quantum
level by
\begin{gather}
{\cal K}_-(\theta;\alpha)=
\begin{pmatrix}
A(\theta) & B(\theta) \\
D(\theta) & E(\theta)
\end{pmatrix}
\end{gather}
and using the expression for the $R$-matrix written above, the
reflection equation \eqref{qKK-} now reduces to eight functional
equations:
\begin{align*}
 (i)~~~~~~&a_-c_+
 \left(B D' -B' D \right)+a_-a_+
 [A ,A']=0,\\
(ii)~~~~~~&b_-b_+
 \big(A E'-E'A \big)+
 c_-c_+
 [E ,E']+\ c_-a_+
 \big(D B' -D' B \big)=0,\\
(iii)~~~~~~&a_-c_+
 \big(D B' -D' B \big)+
  a_-a_+
 [E ,E']=0,\\
(iv)~~~~~~&c_-b_+
 \big(E A'-E'A \big)+ b_-c_+
 \big(A A'-E'E\big)+\ b_-a_+
 \big(B D' -D' B \big)=0,\\
(v)~~~~~~& b_-b_+AD'  +
 c_-c_+ED'+\ c_-a_+DA'
 -\ a_-a_+D' A -\ a_-c_+E'D=0,\\
(vi)~~~~~~& b_-a_+BE' +
 c_-b_+EB'+\ b_-c_+AB'
 -\ a_-b_+E'B=0,
\end{align*}
where we use the shorthand notations $a_-=a(\theta-\theta')$,
$a_+=a(\theta+\theta'-\alpha)$ and similarly for $b$ and $c$ as
well as $A=A(\theta)$ and $A'=A(\theta')$ and similarly for $B, D$
and $E$. The remaining two equations are obtained from $(v)$ and
$(vi)$ through the substitutions $A\leftrightarrow E$ and
$B\leftrightarrow D$. Let us now focus on solutions of kind (III).
To find the solution of the equations $(i)-(vi)$, we assume the
following form for the $K$-matrix:
\beqa A(\theta)&=& Fe^{\theta+\alpha/2} -
Ge^{-\theta-\alpha/2}\qquad \quad \qquad E(\theta)=
Ge^{\theta+\alpha/2} - Fe^{-\theta-\alpha/2}\nonumber \\
B(\theta)&=& -2\sinh^2\theta+2\bp\qquad \qquad \qquad \ D(\theta)=
\ 2\sinh^2\theta-2\bm \nonumber
 \eeqa
Setting $\gamma=\alpha$ in case (III), the generators
$\{F,G,\bp,\bm \}$ have to satisfy the following relations:
\beqa\label{aa}
&& [F,G]=-2\sinh\alpha(\bp-\bm )\ ,\qquad [\bp,\bm
]
=\frac{\sinh 2\alpha}{2}(F^2-G^2)\, ,\nonumber\\
&& F\bm e^{-\alpha}-\bm Fe^{\alpha} - F\sinh\alpha +
G\sinh2\alpha/2=0\, ,\nonumber \\
&& G\bm e^{\alpha}-\bm Ge^{-\alpha} + G\sinh\alpha -
F\sinh2\alpha/2=0\, ,\\
&& F\bp e^{-\alpha}-\bm Fe^{\alpha} - F\sinh\alpha +
G\sinh2\alpha/2=0\, ,\nonumber\\
&& G\bp e^{\alpha}-\bp Ge^{-\alpha} + G\sinh\alpha -
F\sinh2\alpha/2=0\,.\nonumber
\eeqa
One can find a representation for the generators $\{F,G,\bp,\bm
\}$ satisfying these quadratic-linear relations in terms of the
position and momentum operators $q$ and $p$ with commutation
relation
\beqa [p(t),q(t)]=\al. \eeqa
This leads to the following solution of the quantum reflection
equation \eqref{qKK-}:
\begin{equation}
\hspace{-5mm} {\cal K} _{-}(\th;\al)\! = \!
\begin{pmatrix}
\cosh(p-q)e^{\th+\al/2}
-\cosh(p+q)e^{-\th-\al/2} & 2\sinh^2(q)-2\sinh^2(\th)\\
2\sinh^2(\th)-2\sinh^2(p) & \cosh(p+q)e^{\th+\al/2}
-\cosh(p-q)e^{-\th-\al/2}\label{solqK-} \\
\end{pmatrix}.
\end{equation}
%
This ${\cal K}_-(\th;\al)$ satisfies (\ref{qKK-}) also in regime
(I) with $\ga=-\al$. With the substitution $\th\rightarrow i\th$
and $\ga\rightarrow i\ga$ the solution corresponding to the regime
(II) follows immediately. The matrix ${\cal K}_+(\th;\al)={\cal
K}^t_-(-\th;\al)$ provides a solution of \eqref{qKK+}. The
classical reflection matrix \eqref{solK} used in the previous
section is obtained as the classical limit of ${\cal
K}_-(\th;\al)$.

As was observed by Sklyanin, given a c-number valued solution $K$
of the reflection equation one can always construct an infinite
number of additional operator valued solutions by dressing $K$
with $R$-matrices to $RKR$, see \cite{Skl89} for details. The
solution which we derived above is probably not of this
factorizable form.

\section{Discussion\label{sect:d}}
We have explicitly constructed an integrable classical field
theory with extra degrees of freedom living at the boundary. We
have shown how to obtain concrete expressions for the Hamiltonian
and higher spin conserved charges.

While in this paper the sinh-Gordon model coupled to an oscillator
at the boundary was given only as an example of the kind of model
one can obtain, we believe that it is of great interest in its own
right and we intend to study both its classical solutions and its
quantization. In both endeavors we will be helped by the large
amount of work which has been done already on the sine-Gordon
model with fixed boundary conditions, see for example
\cite{Tar91,Gho94,Sal94,Mat00,Baj01}.

Of particular interest are the classical solutions describing
oscillating boundary states. In the case of the fixed boundary
conditions these so called boundary breather solutions were found
and quantized semiclassically to obtain the spectrum of boundary
states \cite{Cor99,Cor00}. In our model we will look for classical
solutions in which not only the field $\phi(x)$ near the boundary
but also the boundary variables $p$ and $q$ oscillate. It is
possible that there will be two or more solutions with the same
energy, they might for example be obtained from one another
through the transformation $\phi\rightarrow -\phi, p\rightarrow
-q, q\rightarrow p$ which is a symmetry of the theory. This
question of how many degenerate states there are is very relevant
to the determination of the reflection matrices. As in the case of
fixed boundary conditions described in \cite{Gho94}, the soliton
reflection matrices will have to satisfy the quantum reflection
equation. However if there is a degeneracy of boundary states then
one will look for operator valued solutions of the reflection
equation so as to be able to describe processes where the
reflection of a soliton rotates the degenerate boundary states
among themselves.

The sinh-Gordon model is just the simplest member of the family of
affine Toda field theories. For all Toda theories there exist
integrable boundary conditions of the form \cite{Cor94}
\begin{equation}
  \partial_x\vec{\phi}(0)=
  \sum_{i=0}^r A_i\vec{\alpha}_i
  e^{\vec{\alpha}_i\cdot\vec{\phi}(0)/2}\,.
\end{equation}
However the boundary parameters $A_i$ were found to be restricted
by the requirement of integrability to only a small discrete set
of values. It would be nice to see if one could replace the fixed
parameters $A_i$ by the dynamical variables of a mechanical system
in the way in which we have done for the sinh-Gordon model in this
paper. The fixed boundary conditions might then arise as the
possible stationary points of the boundary system.

Besides the affine Toda field theories, whose integrability is
based on the trigonometric $R$-matrices, there are many integrable
field theories related to rational $R$-matrices, for example the
non-linear Schr\"odinger model and the principal chiral models.
Finding dynamical solutions of the reflection equation for these
rational models is simpler than in the trigonometric case and many
are already known \cite{Kuz95}. These could be used to construct
integrable couplings to boundary mechanical systems for these
field theories. \vspace{4mm}

\noindent{\bf Acknowledgements:} We would like to thank Chris
Fewster, Evgueni Sklyanin and Anastasia Doikou for very helpful
discussions and Vadim Kuznetsov for showing us his solutions of
the reflection equation. PB is supported by Marie Curie fellowship
HPMF-CT-1999-00094 and GWD by an EPSRC advanced fellowship.

\end{document}